# Soliton Formation in Whispering-Gallery-Mode Resonators via Input Phase Modulation

Hossein Taheri,[1] Ali A. Eftekhar,[1] Kurt Wiesenfeld,[2] Ali Adibi[1,*]

[1]*School of Electrical and Computer Engineering, Georgia Institute of Technology, 777 Atlantic Dr NW, Atlanta, GA 30332-0250, USA*
[2]*School of Physics, Georgia Institute of Technology, 837 State Street, Atlanta, GA 30332-0430, USA*
*Corresponding author: ali.adibi@ece.gatech.edu

We propose a systematic method for soliton formation in whispering-gallery-mode (WGM) resonators through input phase modulation. Our numerical simulations of a variant of the Lugiato-Lefever equation suggest that modulating the input phase at a frequency equal to the resonator free-spectral-range and at modest modulation depths provides a deterministic route towards soliton formation in WGM resonators without undergoing a chaotic phase. We show that the generated solitonic state is sustained when the modulation is turned off adiabatically. Our results support parametric seeding as a powerful means of control, besides input pump power and pump-resonance detuning, over frequency comb generation in WGM resonators. Our findings also help pave the path towards ultra-short pulse formation on a chip.

Starting from a continuous wave (CW) input laser, parametric frequency conversion in optical whispering-gallery-mode (WGM) resonators of high quality factor can give rise to a wide spectrum of equidistant frequency lines known as an optical frequency comb. Frequency combs have been demonstrated in a variety of experimental platforms [1]. They have also been the subject of several theoretical studies. Mathematical models describing the temporal evolution of the individual lines in the spectrum of an optical comb have been proposed [2]. It has also been shown that the spatiotemporal evolution of the total field envelope corresponding to a frequency comb is governed by a damped and driven nonlinear Schrödinger equation, usually referred to as the Lugiato-Lefever equation (LLE) [3-7]. The frequency domain and spatiotemporal pictures have been shown to be equivalent both physically and computationally [8]. Using these models, the existence of stable localized sub-picosecond temporal dissipative solitons has been theoretically predicted and experimentally demonstrated [4,5,9].

Besides paving the path toward compact ultra-short pulse sources, temporal solitons generated from WGM resonators have highly desirable spectral characteristics such as broadband, low-noise frequency spectra and exceptionally small line-to-line amplitude variations. Mode-locking of these frequency combs, however, has proven to be a delicate task. For instance, the solitons demonstrated in [9] were observed while adiabatically sweeping the laser pump frequency detuning from the nearest microcavity resonance. Further investigations have shown that abrupt changes in the detuning or the pump power could also direct the system towards solitonic attractors, and it has been suggested that the system should necessarily go through a chaotic state before achieving a soliton [10,11]. To date, however, no systematic method has been proposed for the mode-locking of microresonator-based frequency combs. This limits the applications of temporal solitons and signifies a challenge towards their chip-scale realization.

In this letter, we propose a scheme for soliton formation based on driving the resonator by a phase-modulated CW laser pump. In essence, this scheme falls under the general heading of parametric seeding which has previously been studied in special cases [12-14]. Parametric seeding provides a means of controlling combs and pulses generated in WGM resonators through the introduction of some seed frequencies and manipulating their amplitudes and, importantly, their phases. We note that a variant of the LLE governs microresonator-based frequency comb generation seeded by input phase modulation. Our numerical simulations of this equation suggest that input phase modulation provides a deterministic path, without having to walk the system through a chaotic phase, towards soliton formation in WGM resonators. We also show that the input modulation can be turned off adiabatically without affecting the generated solitons. Combined with optoelectronic modulators, input phase modulation introduces a viable approach for ultra-short pulse formation on a chip and making the advantages of temporal solitons available at small footprints.

Figure 1 shows the schematic of the structure under study, where an access waveguide is coupled to a WGM resonator. We consider exciting the resonator with a CW laser pump with an amplitude proportional to $\mathcal{F}_0$ and with a frequency $\omega_p$ in the vicinity of a cavity resonance denoted by $\omega_{\eta_0}$. The resonance frequencies of the resonator, assumed to be the different azimuthal orders of the same radial order mode, are centered with respect to this pumped resonance, i.e., each resonance frequency is written as $\omega_{\eta_0} + \omega_\eta$ while its mode number is written as $\eta_0 + \eta$, where $\eta_0$ is the mode number of the pumped resonance and $\eta$ is an integer, $\eta \in \{0, \pm1, \pm2, \cdots\}$. In this notation, $\eta = 0$ and $\omega_0 = 0$ correspond to the pumped (or central) resonance frequency $\omega_{\eta_0}$. The phase of the input laser is modulated, either off-chip or on-chip, at a frequency $\omega_M$ and with a modulation depth $\delta_M$ before coupling into the resonator. The modulation leads to the generation of equidistant sidebands with frequency spacing $\omega_M$ around the pump [15]. These sidebands can be written as

$$\mathcal{F}_\eta = \mathcal{F}_0 J_\eta(\delta_M) \exp\left[i(\sigma + \eta\omega_M - \omega_\eta)t\right] \quad (1)$$

where $\sigma = \omega_p - \omega_{\eta_0}$ is the frequency detuning between the pump and the resonance nearest to it. The function $J_\eta(\delta_M)$ is the Bessel function of the first kind and of order $\eta$ and represents the amplitude of the $\eta$-th sideband of the pump. The exponent in Eq. (1) is the detuning between each modulation-induced sideband frequency $\omega_p + \eta\omega_M$ and its nearest cavity resonance $\omega_{\eta_0} + \omega_\eta$. Using Eq. (1) and following the procedure of [6] with the same assumptions, we find the equation governing the spatiotemporal evolution of the total field envelope $\mathcal{A}(\theta,t)$ (also referred to as the waveform in this letter), i.e.,

$$\frac{\partial \mathcal{A}}{\partial t} = -\left(\frac{1}{2}\Delta\omega_0 + i\sigma\right)\mathcal{A} + (\omega_M - \zeta_1)\frac{\partial \mathcal{A}}{\partial \theta}$$
$$-i\frac{\zeta_2}{2}\frac{\partial^2 \mathcal{A}}{\partial \theta^2} - ig_0|\mathcal{A}|^2\mathcal{A} + \frac{1}{2}\Delta\omega_0 \mathcal{F}_0 e^{-i\delta_M \sin\theta}. \quad (2)$$

Here, $\theta$ is the polar angle measured around the resonator circumference and $t$ represents time. We note that this equation is written in a frame rotating with an angular velocity of magnitude $\omega_M$. The field envelope $\mathcal{A}(\theta,t)$ is normalized such that the spatial integral of $|\mathcal{A}(\theta,t)|^2$ over $\theta$ at each moment is proportional to the total number of photons in the resonator. The parameter $\Delta\omega_0 = \omega_{\eta_0}/Q_L$ is the mode linewidth of the pumped resonance, $Q_L$ being the loaded quality factor at this frequency. The parameters $\zeta_1$ and $\zeta_2$ are the first and second order dispersion coefficients respectively. All coefficients $\zeta_n$ of order $n \geq 3$ have been neglected in this derivation, but could be included in the formalism in a straightforward way [9]. The coefficient $\zeta_1$ is the first-order inter-modal spacing and equals the free-spectral-range (FSR). The parameter $g_0 = n_2 c\hbar\omega_{\eta_0}^2 / n_0^2 V_0$ is the four-wave mixing (FWM) gain at $\omega_{\eta_0}$, where $c$ is the speed of light in vacuum, $\hbar$ is the reduced Planck's constant, $n_2$ is the resonator's Kerr coefficient and $n_0$ is its refractive index at $\omega_{\eta_0}$, and $V_0$ is the effective volume of the pumped mode.

As seen in Eq. (2), the effect of phase modulation at the input is to generate a spatially varying excitation and an effective first-order dispersion term. If the input is modulated with $\omega_M = \zeta_1$, the first-order dispersion term vanishes, and the waveform evolution in time and space will be governed by a variant of the LLE whose input is a function of the spatial coordinate. Due to its practical significance, we will focus on this case in the rest of this letter. Before reporting simulation results, we cast the governing modified LLE in the normalized form [6]

$$\frac{\partial \psi}{\partial \tau} = -(1+i\alpha)\psi - i\frac{\beta}{2}\frac{\partial^2 \psi}{\partial \theta^2} + i|\psi|^2\psi + F e^{i\delta_M \sin\theta}. \quad (3)$$

The field envelope $\mathcal{A}$ and external pump amplitude $F$ have been normalized with respect to the modulus of the comb generation threshold field envelope $|\mathcal{A}_{th}| = (\Delta\omega_0/2g_0)^{1/2}$ [2] such that $\psi = \mathcal{A}^*/|\mathcal{A}_{th}|$ and $F = \mathcal{F}_0^*/|\mathcal{A}_{th}|$. (A star super-index denotes complex conjugation.) The evolution time has been rescaled by twice the pumped resonance

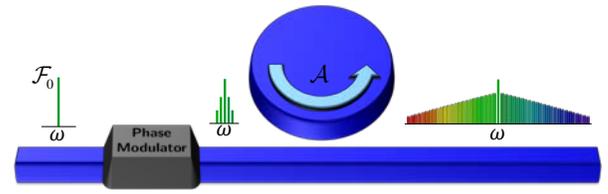

Fig. 1 Schematic of the structure under study. A waveguide is side-coupled to a resonator. The phase of the input laser is modulated before it couples into the resonator. This leads to the generation of sidebands which seed the comb generation process.

photon life-time, i.e. $\tau_0 = 1/\Delta\omega_0$, to give the dimensionless time $\tau = t/2\tau_0$. Finally, $\alpha = -2\sigma/\Delta\omega_0$ and $\beta = -2\zeta_2/\Delta\omega_0$ are the detuning and dispersion coefficients, respectively, both normalized to half the modal linewidth.

The split-step Fourier algorithm [16] has been used for our numerical simulations of Eq. (3). The simulations have been performed for a calcium fluoride (CaF2) resonator of radius $r = 2.5$ mm which has previously been used in both experimental and theoretical investigations [2,12]. For this resonator, $n_0 = 1.43$, $n_2 = 3.2 \times 10^{-20}$ m$^2$/W, and $V_0 = 6.6 \times 10^{-12}$ m$^3$. The dispersion coefficients for this resonator are $\zeta_1 = c/rn_0 = 2\pi \times 13.35$ GHz and $\zeta_2 = 2\pi \times 400$ Hz. The free-space wavelength of the central resonance is $\lambda_{\eta_0} = 2\pi/\omega_{\eta_0} = 1560.5$ nm with a corresponding mode number $\eta_0 = 14350$. The loaded quality factor of the cavity at $\omega_{\eta_0}$ is $Q_L = 3 \times 10^9$ with a mode bandwidth of $\Delta\omega_0 = 2\pi \times 64$ kHz.

If our nonlinear system, with its specific value of $\beta$ and in a proper regime specified by the parameters $\alpha$ and $F$, is provided with a suitable initial condition, it will evolve towards a particular fixed point to produce a desired output. In a regime of parameters where solitonic fixed points exist, starting from a cold cavity (in the presence of vacuum fluctuations) will not necessarily lead to the generation of a soliton, and a particular initial condition (e.g., in the form of a weak pulse) is required [17]. This initial condition can alternatively be supplied by sweeping the pump-resonance detuning or changing the input power thereby walking the system through a chaotic state (as in [9-11]), which can be viewed as providing a pool of different initial conditions. The need for a suitable initial condition poses an obstacle to the realization of stand-alone microresonator-based ultra-short pulse sources. We show here that parametric seeding via the modulation of the input pump phase relaxes this constraint.

Figure 2 illustrates the generation of a soliton starting from a cold cavity when the pump phase is modulated at a depth corresponding to the transfer of ~12% of the pump power to the primary sidebands ($\delta_M = 0.65$). As seen in the left panel in Fig. 2(a), a stable sharply peaked soliton is generated at $\theta = \pi/2$. The spectrum of the pulse is smooth and broadband (Fig. 2(a), middle panel) and the time evolution of the phases of the comb lines (right panel) clearly illustrates mode locking, i.e., the establishment of a fixed relationship between the phases of different comb lines, after the soliton is formed. The waveform and frequency spectrum of this pulse (Figs. 2(b) and (c), respectively) are essentially identical to those of a soliton generated from a weak Gaussian pulse as the initial condition such that if we overlay the corresponding curves on top of each other in one figure, they are indistinguishable to the eye. This suggests that parametric seeding by input phase modulation has walked the system towards the same fixed point as that achieved by a particular initial condition. In Fig. 3(d), a closer look is cast upon the spectra of the pulses generated in the two scenarios in the vicinity of the pumped resonance. The spectrum of the soliton produced by phase modulation (blue curve) and that of the soliton evolved from a suitable non-constant initial condition (red spikes) differ only in the few modal fields close to the pump (taller central spike) and are identical otherwise.

Results presented in Fig. 2 assume that the resonator has zero energy at the onset of the input phase modulation. In an experimental setup, however, the precise synchrony of the pump turn-on time and that of the phase modulator might be difficult. To account for this effect, we have shown in Fig. 3(a) the evolution of the intra-cavity field when the phase modulation is applied to a resonator initially at equilibrium. The equilibrium value of the intra-cavity field $\psi_e$ can be found from Eq. (3) by setting all the derivatives equal to zero [17]. As seen in the left panel of Fig. 3(a), in this case a number of pulses are generated in the resonator. The initial number of pulses depends on the total energy inside the resonator when the modulation is turned on [17, 18]. In the presence of pump phase modulation, each pulse is forced to move towards $\theta = \pi/2$ and merge with the other pulses. The pulse generated at $\theta = -\pi/2$ appears first to be stable but that too starts to move towards $\theta = \pi/2$ after $t = 2$ ms. All of the pulses eventually merge, leaving one stable pulse whose waveform and spectrum are the same as those depicted respectively in Figs. 2(b) and (c). These observations suggest that $\theta = \pm\pi/2$ are fixed points; $\theta = \pi/2$ is stable while $\theta = -\pi/2$ is unstable.

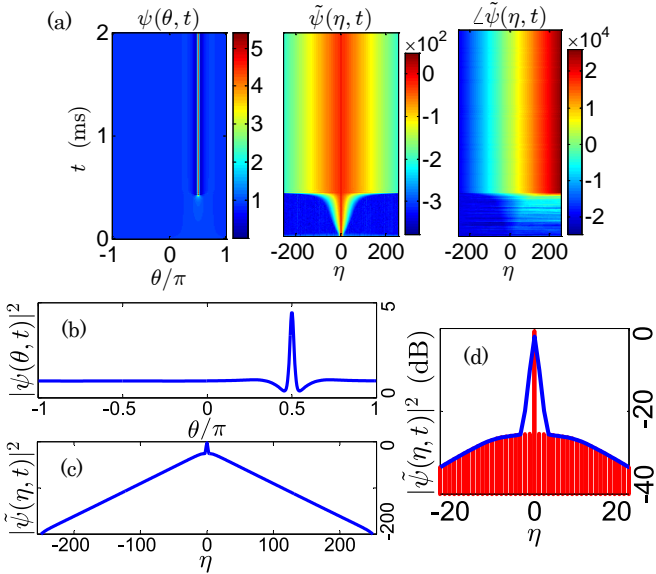

Fig. 2 (Color online) (a) The temporal evolution of the waveform ($\psi$, left), spectrum ($\tilde{\psi}$, middle), and phase ($\angle\tilde{\psi}$, right) of the intra-cavity field starting from zero initial conditions and with input phase modulation. The phase has been unwrapped. (b) The generated pulse corresponding to $t = 2$ ms in (a). (c) The spectrum (in dB) of the pulse corresponding to $t = 2$ ms in (a). (d) Comparison of the spectrum close to the pump for the generated soliton in (a) (blue curve) and a soliton evolved from a weak Gaussian pulse $\psi(\theta, 0) = 0.4 + 0.2 \exp[-\theta^2/2(0.1)^2]$ as the initial condition (red spikes).

The behavior observed in Figs. 2 and 3 can be understood by considering the intra-cavity field momentum define by

$$P = -\frac{i}{2}\int_{-\pi}^{\pi} d\theta \left( \psi^* \frac{\partial \psi}{\partial \theta} - \psi \frac{\partial \psi^*}{\partial \theta} \right). \quad (4)$$

The equation for the momentum time rate of change is

$$\frac{dP}{d\tau} = -2P + 2\delta_M F \int_{-\pi}^{\pi} d\theta \cos\theta \left[ \psi(\theta, \tau)\cos(\delta_M \sin\theta) \right]. \quad (5)$$

In arriving at this equation we have used the fact that a soliton in a system with second-order dispersion has an even spectrum and an odd phase profile and is therefore a real function (see Fig. 2(a)). This could alternatively be seen in light of the hyperbolic secant shape of these solitons [3,4,9,10] since a hyperbolic secant Fourier-transforms into the same functional shape. The first term on the right-hand side of Eq. (5) appears due to the presence of damping while the second term originates from the input phase modulation. The momentum is a decaying exponential function of time in the absence of modulation but will be driven by the second term when modulation is present. The integral in the second term is zero for a soliton centered at $\theta = \pm\pi/2$ since the integrand is an odd function of $\theta$ around either point. It is, however, positive for any pulse $\psi(\theta, \tau)$ centered at $\theta = \theta_0 \in (-\pi/2, \pi/2)$ and is negative for one localized around a point outside of this region in the resonator. Therefore, small perturbations can destabilize a pulse centered at $\theta = -\pi/2$ and move it away from this point while any localized pulse inside the resonator will be dragged away from $\theta = -\pi/2$ and towards $\theta = \pi/2$. In other words, $\theta = \pi/2$ is stable while $\theta = -\pi/2$ is unstable. To further test this explanation, we show in Fig. 3(b) an example where the initial condition in the absence of phase modulation is a weak Gaussian pulse centered at a point slightly to the right of $\theta = -\pi/2$. This pulse evolves rapidly into a soliton. The modulation is turned on at $t = 1$ ms which results in the deflection of the soliton towards $\theta = \pi/2$. The soliton propagates without deviating to either side after reaching this point. The preceding discussion also shows that solitons generated through input phase modulation tend to be more robust than solitons in the absence of this type of seeding, for any perturbation in

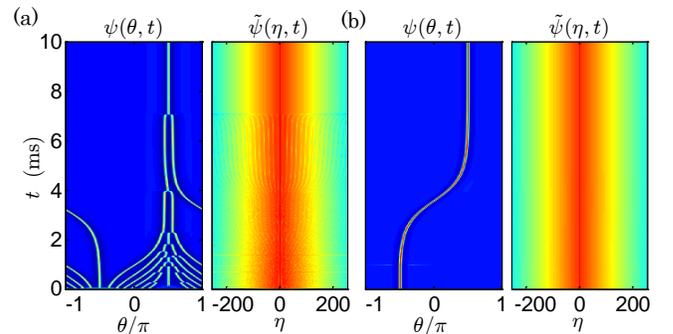

Fig. 3 (a) The temporal evolution of the waveform (left) and spectrum (right) of the intra-cavity field starting from $|\psi(\theta, 0)|^2 = 0.9$ and with input phase modulation. All parameters are the same as those in Fig. 2(a). (b) Same as (a) but starting from a weak Gaussian pulse $\psi(\theta, 0) = 0.2 + \exp[-(\theta + 0.49\pi)^2/2(0.1)^2]$ and without pump phase modulation. The modulation is then turned on at $t = 1$ ms. The vertical axis is the same as that in (a).

the position or the shape of the pulse will be opposed and suppressed by the modulated pump. We note that a non-zero phase for the modulator merely shifts the equilibria.

We indicated earlier that the modulation of the input phase can lead the system towards the same fixed points available in the absence of the modulation and through a suitable input (Fig. 2(d)). It is therefore reasonable to expect that the generated solitons will be sustained if the modulation is turned off adiabatically. This is seen in Fig. 4(a) where a soliton is initially formed by pump phase modulation starting from a cold cavity. From $t = 1$ ms, we slowly reduce the modulation depth $\delta_M$ to zero. The soliton which was originally formed by phase modulation is sustained after the modulation is removed. For the purpose of comparison, we show in Fig. 4(c) the effect of the abrupt removal of the modulation at $t = 2$ ms. A shock is inflicted on the system which leads to the generation of four other pulses. As time passes, the extra pulses on each side of the original soliton move slightly away from each other and stabilize by $t = 4$ ms. As seen in Fig. 4(d), the final spectrum in this case does no longer consist of a comb with smooth envelope and small line-to-line amplitude variations. Note that in Fig. 4 the origin of $\theta$ has been shifted to $\theta = -\pi/2$ for better visualization.

From the perspective of nonlinear dynamics, the vanishing of the parameter $\delta_M$ alters the topography of the equilibria in the system described by Eq. (3). Intuitively, two scenarios are possible when a fixed point of interest remains an attractor after the change is applied. If the change is applied gradually, the state of the system remains in the basin of attraction of the fixed point (Fig. 4(b)).

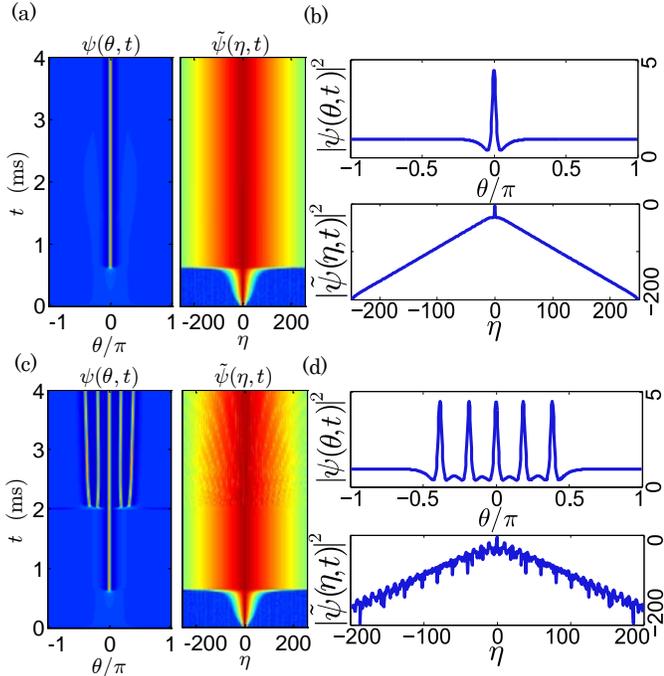

Fig. 4 (Color online) Comparison of the adiabatic and abrupt removal of the input phase modulation. (a) The time evolution of the field envelope (left) and spectrum (right) when the modulation is adiabatically turned off: $\delta_M$ ($t > t_0$) = 0.65 exp[-($t$ - $t_0$)$^2$/2$\sigma_M^2$], $t_0 = 1$ ms, $\sigma_M = 0.6$ ms. (b) Final waveform (top) and spectrum (in dB, bottom) in (a). (c) Same as (a), but for abrupt modulation removal at $t = 2$ ms. (d) Final waveform (top) and spectrum (in dB, bottom) in (c). The field envelope has been shifted for better visualization.

If, however, the change is abrupt, it may place the system in the basin of attraction of a different fixed point and, therefore, the final state of the system will be different (Fig. 4(d)). This intuitive result has been verified rigorously and it has been shown that when a system starts sufficiently close to a stable equilibrium or limit cycle, it will follow this attractor when the parameter ($\delta_M$ in our problem) is varied adiabatically [19]. We note that abrupt removal of the modulation does not necessarily lead to losing the single soliton. In particular, our numerical simulations show that for smaller pump amplitudes, (e.g. $F = 1.3722$) the modulation could safely vanish abruptly.

To conclude, we proposed a method for soliton formation in WGM resonators in a deterministic way through input phase modulation. Using a variant of the LLE, we showed, both numerically and analytically, that parametric seeding by pump phase modulation allows us to manipulate pulses in a WGM resonator and leads to more stable solitons. We also showed that the seeding agent can be removed without affecting the generated solitons. Our findings support parametric seeding as a powerful means of control over frequency comb generation in WGM resonators and help pave the path towards chip-scale ultra-short pulse sources.

We thank Lute Maleki and Andrey B. Matsko from OEwaves Inc. for useful comments on the manuscript. This work was supported by AFOSR grant No. 2106CPZ (G. Pomrenke).